\documentclass[aps,pra,preprint,groupedaddress,showpacs]{revtex4}
\begin{document}

\title{Berry effect in acoustical polarization transport in phononic crystals}

\author{Reza Torabi, Mohammad Mehrafarin}
\affiliation{Physics Department, Amirkabir University of Technology, Tehran 15914,
Iran}
\email{mehrafar@aut.ac.ir}

\begin{abstract}
We derive the semiclassical equations of motion of a transverse acoustical wave packet propagating in a phononic crystal subject to slowly varying perturbations. The formalism gives rise to Berry effect terms in the equations of motion, manifested as the Rytov polarization rotation law and the polarization-dependent Hall effect. We show that the formalism is also applicable to the case of non-periodic inhomogeneous media, yielding explicit expressions for the Berry effect terms.
\end{abstract}


\maketitle

\section{Introduction}
The well-known analogy between the (linearized) equations of elasticity and Maxwell equations (see e.g. \cite {Auld}) gives the advantage of predicting the acoustical analogue of many optical phenomena. In particular, polarization phenomena in optics find their counterparts in transverse acoustic waves. For example, the Rytov polarization rotation law \cite{Rytov} and the polarization dependent Hall effect in optics, which are manifestations of the Berry effect for photons \cite{Tomita,Chiao,Haldane,Berry,Segert,Liberman,Bliokh,Bliokh2,Onoda,Onoda2,Duval}, pertain to transverse acoustic waves as well \cite{Segert,Karal,Bliokh3}. The study of `spin-orbit' interactions, that is the influence of intrinsic degrees of freedom (polarization/spin) on transport properties, has recently formed the subject of spin transport. Because of the dependence of transport on polarization/spin, particles in different polarization/spin states propagate along slightly different (semiclassical) trajectories. 
Examples of spin transport phenomena are the anomalous and the spin Hall effects in solids
\cite{Culcer,Murakami,Sinova,Kato}, analogous effects for relativistic electrons \cite{Berard}, the optical \cite{Bliokh,Bliokh2,Onoda,Onoda2,Duval} and the acoustical \cite{Bliokh3} polarization-dependent Hall effects.

Recently, optical polarization transport in inhomogeneous media, in particular in photonic crystals, has been studied via the semiclassical wave packet approximation \cite{Onoda,Onoda2,Sawada}. The wave packet approximation had proved to be very successful in the study of electronic spin transport in solids, for which it was originally developed \cite{Culcer,Sundaram,Chang}. Here, via the same approach, we obtain analogous results for acoustical polarization transport in a phononic crystal (formed by the periodic variation of the density of the medium), subject to perturbations that vary slowly in space. Phononic crystals are novel materials that owe their name to the analogous effects of photonic crystals on light - they permit the passage of waves only at certain frequencies and can be, therefore, employed to control and confine sound \cite{Sigalas,Trigo,Yang,Zhang,Gorishnyy}. A complication that arises in acoustical polarization transport in inhomogeneous media is the coupling between the transverse and longitudinal waves (here caused by density variations), which means that polarization cannot be maintained. However, the coupling vanishes in the adiabatic limit of slow density variations, rendering polarization an adiabatic invariant. The formalism gives rise to a Berry phase term in the polarization evolution equation, describing the rotation of the polarization vector (the Rytov law), and a Berry curvature term in the equation of trajectory, deflecting the phonons depending on their polarization (the polarization dependent Hall effect). Such Berry effects have been previously established via a {\it different} semiclassical approach (the short-wave or post-geometric approximation) for transverse acoustical transport in {\it non-periodic} inhomogeneous media \cite{Bliokh3}. 

\section{Acoustical Bloch waves}
The dynamics of the displacement vector filed ${\bf U}({\bf x},t)$ in an elastic medium is governed by (see e.g. \cite{Landau})
\begin{equation}
\rho \partial_t^2 {U_i } =
\partial_j \sigma _{ij}  \label{1}
\end{equation}
(summation convention implied) where $\rho ({\bf x})$ is the density of the medium and
$\sigma_{ij}$ is the stress tensor. In an isotropic medium the latter takes the form
$$
\sigma_{ij} =\lambda \delta _{ij}\ \partial_k U_k +\mu (\partial_i U_j+\partial_j U_i)
$$
$\lambda $ and $\mu $ being the Lam\'{e} coefficients (assumed constant throughout). Let us, thus, rewrite equation (\ref {1}) as 
\begin{equation}
\partial_t^2 {\bf U} =\frac{\lambda+2\mu}{\rho}\ \nabla (\nabla \cdot {\bf U}) -\frac{\mu}{\rho}\ \nabla \times
(\nabla \times {\bf U}). \label{2}
\end{equation}
We decompose ${\bf U}$ into a solenoidal part, ${\bf U}_\bot$, and an irrotational part, ${\bf U}_\|$, according to ${\bf U}={\bf U}_\bot+{\bf U}_\|$ (such a decomposition is, of course, always possible for an arbitrary vector). Since the volume change in a deformation is given by $\nabla \cdot {\bf U}$, the solenoidal part of the displacement field does not involve any change in volume as it propagates in the body. It is thus termed the transverse displacement field, which has two independent components (polarization modes). The irrotational part, however, involves compressions and expansions in the body and is termed the longitudinal displacement field, having a single polarization mode. Thence, by taking the curl and the divergence of (\ref{2}), we obtain
\begin{eqnarray}
\nabla \times (\partial _t^2 {\bf U}_\bot - v_\bot^2\ \nabla^2 {\bf U}_\bot)=-\nabla \times (\partial _t^2 {\bf U}_\| - v_\|^2\ \nabla^2 {\bf U}_\|)= \nabla 
v_\|^2 \times \nabla^2  {\bf U}_\| \nonumber \\
\nabla \cdot (\partial _t^2 {\bf U}_\bot - v_\bot^2\ \nabla^2 {\bf U}_\bot)=-\nabla \cdot (\partial _t^2 {\bf U}_\| - v_\|^2\ \nabla^2 {\bf U}_\|)= -\nabla v_\bot^2 \cdot \nabla^2 {\bf U}_\bot \label{3}
\end{eqnarray}
where
$$v_\bot ({\bf x})=\sqrt {\frac {\mu}{\rho({\bf x})}},\ \ \ \ v_\|
({\bf x})=\sqrt {\frac{\lambda +2\mu}{\rho({\bf x})}}
$$ 
The curl and the divergence of the expressions in parentheses would vanish everywhere if the density were uniform, implying that the expressions must be zero identically, i.e., 
\begin{eqnarray}
\partial_t^2{\bf U}_\bot-v_\bot^2
\nabla ^2{\bf U}_\bot =0,\ \ \ \ 
\partial_t^2{\bf U}_\|-v_\|^2
\nabla ^2{\bf U}_\| =0 \label{4}
\end{eqnarray}
The transverse and longitudinal modes would, thus, propagate independently as waves with speeds $v_\bot$ and $v_\|$, respectively. In presence of density variations, however, equations (\ref{3}) are coupled. Nevertheless, introducing a slow (adiabatic) variation of the density, the right hand sides of (\ref{3}) may be neglected and the equations still decouple. This decoupling is vital in acoustical polarization transport, because otherwise, polarization cannot be maintained (to be contrasted with optics). The adiabatic approximation eliminates the coupling between the transverse and longitudinal modes and renders polarization an adiabatic invariant. Therefore, in the adiabatic limit, (\ref{4}) still applies where $v_\bot$ and $v_\|$ are now local wave speeds, and we shall be interested only in the transverse wave equation, of course. 

The transverse index of refraction, $n_\bot$, of an inhomogeneous medium is defined by $v_\bot ({\bf x})=c_\bot/ n_\bot ({\bf x})$, where $ c_\bot$ is the transverse wave speed in a reference homogeneous medium. For convenience, we shall take $c_\bot=1$. Introducing the monochromatic wave solution 
$$
{\bf U}_\bot({\bf x},t)= \sum_{\lambda=1,2} {\bf e}_\lambda U_\lambda({\bf x},t)
= \sum_\lambda {\bf e}_\lambda 
u_\lambda({\bf x})e^{-i\omega t} 
$$
${\bf e}_\lambda$ being the (orthonormal) transverse polarization vectors, into the transverse wave equation yields the eigenvalue equation 
\begin{equation}
-\frac{1}{n_\bot^2({\bf x})}\nabla^2 u_\lambda =\omega^2 u_\lambda \label{5}
\end{equation} 
for polarized transverse waves. The differential operator in (\ref{5}) is not Hermitian. However, by introducing the Hermitian operator, 
$$
\hat{Q}=-\frac{1}{n_\bot({\bf x})}\nabla^2 \frac{1}{n_\bot({\bf x})}
$$
the equation can be written in the self-adjoint form
\begin{equation}
\hat{Q} \psi_\lambda=\omega^2 \psi_\lambda \label{6}
\end{equation}
where
$$
\psi_\lambda({\bf x})=n_\bot({\bf x}) u_\lambda({\bf x}).
$$
Thus, with the weighting function $n_\bot$, eigenfunctions $u_\lambda$ form a complete orthogonal set that is quasi (adiabatically) transverse. In a phononic crystal defined by imposing periodic condition on the refractive index through $\rho({\bf x})$, the eigenfunctions of (\ref{6}) are Bloch functions (Bloch's theorem), viz.,
\begin{equation} 
\psi_{\lambda n} ({\bf x},{\bf q})= e^{i{\bf q}.{\bf x}} w_{\lambda n} ({\bf x},{\bf q}) \label{7} 
\end{equation}
where the corresponding eigenfrequencies, $\omega_n({\bf q})$, exhibit band structure as a result of the periodic condition. Here, ${\bf q}$ is the lattice momentum (restricted to the first Brillouin zone), $n$ is the band index and $w_{\lambda n} ({\bf x},{\bf q})$ is periodic in ${\bf x}$ with the periodicity of the crystal lattice (viz. $\rho$). The acoustical Bloch waves (\ref{7}) form a complete orthonormal set,
$$
\int d{\bf x}\ \psi_{\lambda n}^\ast ({\bf x},{\bf q})\psi_{\lambda^\prime n^\prime}({\bf x},{\bf q}^\prime)= \delta_{\lambda \lambda^\prime} \delta_{n n^\prime} \delta({\bf q}-{\bf q}^\prime) 
$$
and, thus, serve as a convenient basis for constructing phononic wave packets. In the following, we ignore the possibility of interband transitions and consider a single frequency band at a time, so that reference to the band index may be dropped. 

\section{Wave packet of Bloch phonons}

Let us proceed to the quantum level and introduce the ket notation, $\psi_\lambda ({\bf x},{\bf q})=\langle {\bf x} |\psi_\lambda ({\bf q}) \rangle$, $w_\lambda ({\bf x},{\bf q})=\langle {\bf x} |w_\lambda ({\bf q}) \rangle$, etc. We consider a wave packet $|\Psi \rangle$ of Bloch phonons with a momentum ${\bf k}$ that is well defined on the scale of the Brillouin zone. Such a wave packet must be spread in real space over many primitive cells about a mean value ${\bf r}$. In the semiclassical approximation, which describes the wave packet on a spatial scale much greater than its spread, the location of the packet may be specified by ${\bf r}$. The semiclassical approach, thus, associates with each wave packet a position ${\bf r}$ and a momentum ${\bf k}$ (as well as a polarization state, see shortly), and predicts how these parameters evolve as the wave propagates in the medium. In the course of propagation, the wave packet may generally experience perturbations in its neighborhood (such as deformation strain fields) that vary slowly within its width. Such slowly varying perturbations may be treated classically provided they are evaluated at the wave packet's mean position ${\bf r}$. The semiclassical Hamiltonian $\hat{H}$, which determines the dynamics of the wave packet via the time-dependent Schr\"{o}dinger equation $i|\dot{\Psi}\rangle=\hat{H}|\Psi\rangle$ ($\hbar=1$), is then the sum of a periodic (quantum mechanical) Hamiltonian and the classically treated perturbations. Thus, $\hat{H}$  has the periodicity of the unperturbed Hamiltonian and involves ${\bf r}$ as a parameter. Whence ${\bf r}$ enters in the (Bloch) eigenstates and the corresponding eigenfrequencies parametrically. 

We, thus, construct the (normalized) wave packet according to 
\begin{equation}
|\Psi(t) \rangle =\sum_\lambda z_\lambda(t) \int_{BZ} d{\bf q}\ g({\bf q},t) |\psi_\lambda({\bf r},{\bf q}) \rangle \label{8}
\end{equation}
where $g({\bf q},t)$ and $z_\lambda(t)$ are amplitudes with the normalizations
\begin{eqnarray}
\int_{BZ} {d{\bf q} | {g({\bf q},t)} |} ^2=1 \nonumber \\
\sum_\lambda  {| {z_\lambda(t)} |^2=1}. \nonumber
\end{eqnarray}
The momentum distribution $|g|^2$ is assumed to be sharply peaked around its mean value ${\bf k}$, 
$$
{\bf k}(t)=\int_{BZ} d{\bf q}\ |g|^2{\bf q}
$$
which is regarded as the momentum of the wave packet. In the complex two-dimensional polarization space, the polarization state of the wave packet is represented by the parameters $z_\lambda$. The wave packet $|\Psi(t)\rangle \equiv |\Psi(\vec {R}(t))\rangle$ is, thus, characterized by the parameters $\vec {R}=(R_a) \equiv ({\bf r}, {\bf k}, z_\lambda)$. (For the sake of clarity, we reserve boldface letters for three-dimensional vectors, and middle alphabet indices for their components.) The location of the wave packet, which is specified by its mean position $\langle \Psi |\hat{\bf x} | \Psi \rangle$, is self-consistently given by
\begin{equation}
{\bf r}(t) =\int_{BZ} d{\bf q}\ |g|^2
\left( \nabla _{\bf q} S+
\langle W
|i\nabla_{\bf q} | W\rangle \right), \label{r}
\end{equation}
where $S({\bf q},t)$ is the phase of the momentum amplitude, viz. $g=|g|e^{-iS}$, and $|W({\bf r},{\bf q}, z_\lambda) \rangle=\sum_\lambda z_\lambda |w_\lambda({\bf r},{\bf q}) \rangle$ ($\langle W |W\rangle =1$). In view of our assumption of a well localized wave packet in the momentum space, (\ref{r}) reduces to
\begin{equation}
{\bf r}(t)\approx {\bf \nabla }_{\bf k} S({\bf k},t)+ \langle W(\vec {R})|i \nabla_{\bf k}|W(\vec {R}) \rangle. \label{10}
\end{equation}

\section{Dynamics of the wave packet}
The dynamics of the wave packet is determined by the time-dependent Schr\"{o}dinger equation which is derivable from the Lagrangian 
$$
{\cal L}(|\Psi \rangle, |\dot{\Psi} \rangle,t)=i\langle \Psi | \dot{\Psi}\rangle-\langle \Psi |\hat{H}| \Psi \rangle
$$
via the Euler-Lagrange equation \cite{Jackiw}. Thus, using (\ref{8}),
\begin{eqnarray}
{\cal L}= \int_{BZ} d{\bf q}\ |g|^2 \left( i\langle W |\dot{W} \rangle 
+\partial_t S-\omega \right) 
\approx \sum_\lambda \dot{z}_\lambda \langle W(\vec {R})|i \partial_{z_\lambda} |W(\vec {R})\rangle 
 +\dot{\bf r} \cdot \langle W(\vec {R})|i \nabla_{\bf r}|W(\vec {R}) \rangle \nonumber\\
+\partial_t S({\bf k},t)- \omega({\bf r},{\bf k} )\nonumber
\end{eqnarray}
Since,
$$
\partial_t S({\bf k},t)= \dot{S}({\bf k},t)-\dot{{\bf k}} \cdot \nabla_{\bf k} S({\bf k},t) \approx \dot{S}-\dot{{\bf k}} \cdot \left({\bf r}-\langle W(\vec {R})|i \nabla_{\bf k}|W(\vec {R}) \rangle \right)
$$
having used (\ref{10}) in the last expression, the Lagrangian becomes (up to an inconsequential total time derivative $\dot{S}$),
\begin{equation}
{\cal L}({\bf R},\dot{\bf R},t)\approx  \dot{\vec {R}} \cdot \vec {A}(\vec {R})-\dot{\bf k} \cdot {\bf r}- \omega({\bf r},{\bf k}), \label{lag}
\end{equation}
where 
$$
\vec {A}(\vec {R})=\langle W(\vec {R})
|i\nabla_{\vec {R}} | W(\vec {R}) \rangle
$$
is the Berry connection (gauge potential) in the parameter space. It has components
\begin{eqnarray}
A_{r_i}=\langle W(\vec {R})
|i\partial_{r_i} | W(\vec {R}) \rangle= \sum_{\lambda \lambda \prime} \langle w_\lambda ({\bf r},{\bf k})|i\partial_{r_i}| w_{\lambda \prime} ({\bf r},{\bf k}) \rangle z^\ast_\lambda z_{\lambda \prime} \nonumber\\
A_{k_i}=\langle W(\vec {R})
|i\partial_{k_i} | W(\vec {R}) \rangle= \sum_{\lambda \lambda \prime} \langle w_\lambda ({\bf r},{\bf k})|i\partial_{k_i}| w_{\lambda \prime} ({\bf r},{\bf k}) \rangle z^\ast_\lambda z_{\lambda \prime} \nonumber\\
A_{z_\lambda}=\langle W(\vec {R})
|i\partial_{z_\lambda} | W(\vec {R}) \rangle= i z_\lambda^\ast\ \ \ \ \ \ \ \ \ \ \ \ \ \ \ \ \ \nonumber
\end{eqnarray}
The Berry curvature tensor (gauge field) associated with the Berry connection has components
$$
\Omega _{ab}(\vec {R})=\partial_{R_a} A_b(\vec {R})- \partial_{R_b} A_a(\vec {R}).
$$
Clearly, only $A_{r_i}$ and $A_{k_i}$ give rise to non-vanishing components which are $\Omega_{r_i r_j}=(\nabla _{\bf r} \times {\bf A}_{\bf r})_k \equiv ({\bf \Omega}_{\bf r})_k$, $\Omega_{k_i k_j}=(\nabla _{\bf k} \times {\bf A}_{\bf k})_k\equiv ({\bf \Omega}_{\bf k})_k$ ($i,j,k$ in cyclic permutations) and $\Omega_{r_i k_j}$. By introducing matrix notation in the polarization space, these connections can be expressed more conveniently as 
$$
A_{r_i}=\langle z| \hat{A}_{r_i}| z\rangle ,\ \ \ A_{k_i}=\langle z| \hat{A}_{k_i}| z\rangle
$$
where $|z \rangle$ is the column matrix with elements $z_\lambda$ representing the polarization state ($\langle z| z\rangle=1$), $\hat{A}_{r_i}({\bf r},{\bf k})$ is the (Hermitian) matrix with elements $\langle w_\lambda({\bf 
r},{\bf k}) |i\partial_{r_i}| w_{\lambda \prime}({\bf r},{\bf k}) \rangle$ and likewise for $\hat{A}_{k_i}({\bf r},{\bf k})$. 
The Lagrangian (\ref {lag}) can be, thus, expressed as
\begin{equation}
{\cal L}\approx i \langle z| \dot{z}\rangle+\dot{\bf r} \cdot {\bf A}_{\bf r}+\dot{\bf k} \cdot {\bf A}_{\bf k} -\dot{\bf k} \cdot {\bf r}- \omega. \label{12}
\end{equation}

Effective Lagrangians like (\ref{12}), which determine the semiclassical dynamics of the wave packets, are central to the semiclassical wave packet approximation. The equations of motion that follow from (\ref{12}) are
\begin{eqnarray}
|\dot {z}\rangle =i(\dot{\bf r} \cdot \hat{\bf A}_{\bf r}+\dot{\bf k} \cdot \hat{\bf A}_{\bf k})|z\rangle \label{1st}\\
\dot{r}_i = ({\bf v}_g + \dot{\bf k} \times {\bf \Omega}_{\bf k})_i-\Omega_{k_i r_j} \dot{r}_j \label{2nd}\\
\dot {k}_i =(- \nabla_{\bf r} \omega+ \dot{\bf r} \times {\bf \Omega}_{\bf r})_i +  \Omega_{r_i k_j}\dot{k}_j  \label{3rd}
\end{eqnarray}
where ${\bf v}_g({\bf r},{\bf k})=\nabla_{\bf k} \omega$ is the group velocity of the wave packet. These are identical in form to the equations obtained in \cite{Sawada} for optical polarization transport in photonic crystals. Equation (\ref{1st}), which describes the time evolution of
the polarization, has the solution $|z(t)\rangle=e^{i\hat{\gamma}} |z(0)\rangle$, where $\hat{\gamma}=\int_C (d{\bf r} \cdot \hat{\bf A}_{\bf r}+ d{\bf k} \cdot \hat{\bf A}_{\bf k})$ is the (Hermitian) Berry-phase matrix associated with the wave packet trajectory $C$ in parameter space. The phase shift includes the contribution of the directional change of propagation (second term in the integrand) and thus generalizes the Rytov polarization rotation law. In the remaining two equations, the first terms on the right-hand side reproduce geometrical acoustics. The second term in (\ref{2nd}) represents an anomalous velocity, which, being polarization dependent, gives rise to an acoustical polarization-dependent Hall effect, just as in the optical counterpart \cite{Onoda,Onoda2,Sawada}. The remaining terms owe to the existence of slowly varying perturbations. They are the second term in (\ref{3rd}), which corresponds to a Lorentz force, and the third terms embodying the effect of mixed curvatures.

\section{Non-periodic inhomogeneous media}
The above formalism is immediately applicable, just as in the optical counterpart, to the case of slightly inhomogeneous media with {\it non-periodic} refractive index, too. This, in the absence of perturbations, presents an example where the Berry connection/curvature can be calculated explicitly, reproducing the corresponding results of \cite{Onoda,Onoda2} for photons. In this case, a basis for the wave packet expansion is provided by $|\psi_\lambda({\bf q}) \rangle=e^{i{\bf q} \cdot {\bf x}} |{\bf e}_\lambda({\bf q}) \rangle$ so that $|w_\lambda\rangle \rightarrow |{\bf e}_\lambda \rangle$, ${\bf e}_\lambda$ being the polarization vectors. The latter can be written in the helicity basis as ${\bf e}_\lambda= \frac{1}{\sqrt{2}} ({\bf i}_\theta \pm i {\bf i}_\phi)$, ${\bf i}_\theta$ (${\bf i}_\phi$) being the zenithal (azimuthal) unit vector in the spherical coordinates of the momentum space. Thence,
 $$\hat{\bf A}_{\bf r}({\bf k})=0,\ \ \ \hat{\bf A}_{\bf k}({\bf k})=\frac{\cot \theta}{k}\hat{\sigma}_3{\bf i}_\phi
$$
where $\hat{\sigma}_3$ is the Pauli matrix. This yields the Berry phase matrix $\hat{\gamma}=\hat{\sigma}_3 \int_C dk_\phi \frac{\cot \theta}{k}$, so that right/left circularly polarized waves experience a geometric phase shift of $\pm \int_C dk_\phi \frac{\cot \theta}{k}$, respectively. The non-vanishing curvature is given by 
$${\bf \Omega }_{\bf k}({\bf k}) =-\frac{{\bf k}}{k^3} \langle z| \hat{\sigma}_3|z \rangle=-\frac{{\bf k}}{k^3} \langle z(0)| \hat{\sigma}_3|z(0) \rangle
$$
where the final expression follows because $\hat{\gamma}$ commutes with $\hat{\sigma}_3$.  Hence, ${\bf \Omega }_{\bf k}({\bf k}) =\mp \frac{{\bf k}}{k^3}$ for right/left circularly polarized waves respectively, which, according to (\ref{2nd}), therefore propagate along different trajectories. These provide explicit examples of the Rytov law and the polarization-dependent Hall effect for transverse phonons. They coincide with the results of \cite{Bliokh3} obtained via the post-geometric approximation for non-periodic inhomogeneous media.


\begin{thebibliography}{widest-label}
\bibitem{Auld} B. A. Auld, {\it Acoustic fields and waves in solids} (John Wiley \& Sons, New York, 1973; Krieger Pub. Co., 1990).
\bibitem{Rytov} S. M. Rytov, Dokl. Akad. Nauk. SSSR {\bf 18}, 263 (1938).
\bibitem{Tomita} A. Tomita and R. Y. Chiao, Phys. Rev. Lett. {\bf 57}, 937 (1986).
\bibitem{Chiao} R. Y. Chiao and Y. S. Wu, Phys. Rev. Lett. {\bf 57}, 933 (1986).
\bibitem{Haldane} F. D. M. Haldane, Opt. Lett. {\bf 11}, 730 (1986).
\bibitem{Berry} M. V. Berry, Nature {\bf 326}, 277 (1987).
\bibitem{Segert} J. Segert, Phys. Rev. A {\bf 36}, 10 (1987).
\bibitem {Liberman} V. S. Liberman and B. Ya. Zel'dovich, Phys. Rev. A {\bf 46}, 5199 (1992).
\bibitem{Bliokh} K. Yu. Bliokh and Yu. P. Bliokh, Phys. Rev. E {\bf 70}, 026605 (2004).
\bibitem{Bliokh2} K. Yu. Bliokh and V. D. Freilikher, Phys. Rev. B {\bf 72}, 035108 (2005).
\bibitem {Onoda} M. Onoda, S. Murakami and N. Nagaosa, Phys. Rev. Lett. {\bf 93}, 083901 (2004).
\bibitem{Onoda2} M. Onoda, S. Murakami and N. Nagaosa, Phys. Rev. E {\bf 74}, 066610 (2006).
\bibitem{Duval} C. Duval, Z. Horv\'{a}th and P. A. Horv\'{a}thy, Phys. Rev. D {\bf 74}, 021701(R) (2006).
\bibitem{Karal} F. C. Karal and J. B. Keller, J. Acoust. Soc. Am. {\bf 31}, 694 (1958).
\bibitem{Bliokh3} K. Yu. Bliokh and V. D. Freilikher, Phys. Rev. B {\bf 74}, 174302 (2006).
\bibitem{Culcer} D. Culcer, A. MacDonald and Q. Niu, Phys. Rev. B {\bf 68}, 045327 (2003).
\bibitem{Murakami} S. Murakami, N. Nagaosa and S. C. Zhang, Science {\bf 301}, 1348 (2003).
\bibitem{Sinova} J. Sinova, D. Culcer, Q. Niu, N. A. Sinitsyn, T. Jungwirth and A. H. MacDonald, Phys. Rev. Lett. {\bf 92}, 126603 (2004).
\bibitem{Kato} Y. K. Kato, R. C. Myers, A. C. Gossard and D. D. Awschalom, Science {\bf 306}, 1910 (2004).
\bibitem{Berard} A. Berard and H. Mohrbach, Phys. Lett. A {\bf 352}, 190 (2006).
\bibitem{Sawada} K. Sawada, S. Murakami and N. Nagaosa, Phys. Rev. Lett. {\bf 96}, 154802 (2006).
\bibitem{Chang} M. C. Chang and Q. Niu, Phys. Rev. B {\bf 53}, 7010 (1996).
\bibitem{Sundaram} G. Sundaram and Q. Niu, Phys. Rev. B {\bf 59}, 14915 (1999).
\bibitem{Sigalas} M. M. Sigalas and E. N. Economou, Solid State Commun. {\bf 86}, 141 (1993).
\bibitem{Trigo} M. Trigo, A. Bruchhausen, A. Fainstein, B. Jusserand and V. Thierry-Mieg, Phys. Rev. Lett. {\bf 89}, 227402 (2002).
\bibitem{Yang} S. Yang, J. H. Page, Z. Liu, M. L. Cowan, C. T. Chan and P. Sheng, Phys. Rev. Lett. {\bf 93}, 024301 (2004).
\bibitem{Zhang} Y. Zhang and Z. Liu, App. Phys. Lett. {\bf 85}, 341 (2004).
\bibitem{Gorishnyy} T. Gorishnyy, C. K. Ullal, M. Maldovan, G. Fytas and E. L. Thomas, Phys. Rev. Lett. {\bf 94}, 115501 (2005).
\bibitem{Landau} L. D. Landau and E. M. Lifshits, {\it Theory of elasticity} (Pergamon Press, Oxford, 1986).
\bibitem{Jackiw}R. Jackiw and A. Kerman, Phys. Lett. A {\bf 71}, 158 (1979).
\end{thebibliography}
\end{document}